\documentclass[twocolumn,10pt, journal, final]{IEEEtran}

\usepackage{epsfig, makeidx,mathtools, color,bigints,amsbsy,amsmath,amssymb,amsfonts,euscript,verbatim,tikz}%
\usetikzlibrary{arrows,snakes,backgrounds,positioning,graphs,trees}
\usepackage{graphicx}
\usepackage{dsfont}
\usepackage{cite}
\usepackage{array}
\usepackage{times}
\usepackage{stfloats}
\usepackage{yhmath}
\usepackage{graphics}
\usepackage{textcomp}
\usepackage{amscd}
\usepackage{psfrag}
\usepackage{rotating}
\usepackage{url}
\usepackage{float}
\usepackage{balance}

\makeatother

\DeclareMathOperator{\nullspace}{null}
\DeclareMathOperator{\spanspace}{span}

\newtheorem{prop}{Proposition}
\newtheorem{definition}{Definition}

\newtheorem{corollary}{Corollary}
\newtheorem{example}{Example}
\newtheorem{theorem}{Theorem}

\newcommand{\R}{\mathbf{R}} 
\newcommand{\C}{\mathbf{C}} 

\begin{document}

\title{Perfect Secrecy in Physical Layer Network Coding Systems from Structured Interference}

\author{David A. Karpuk, \IEEEmembership{Member, IEEE,} Arsenia Chorti, \IEEEmembership{Member, IEEE}
\thanks{D. Karpuk is with the Department of Mathematics and Systems Analysis, Aalto University, Espoo, Finland.  A. Chorti is with the Department of Computer Science and Electronic Engineering, University of Essex, Colchester, United Kingdom.  emails: \{david.karpuk@aalto.fi, achorti@essex.ac.uk\}}
\thanks{D. Karpuk is supported by Academy of Finland Postdoctoral grant $268364$.}
}
\maketitle



\begin{abstract}
Physical layer network coding (PNC) has been proposed for next generation networks. In this contribution, we investigate PNC schemes with embedded perfect secrecy by exploiting structured interference in relay networks with two users and a single relay.  In a practical scenario where both users employ finite and uniform signal input distributions we propose upper bounds (UBs) on the achievable perfect secrecy rates and make these explicit when PAM modems are used.  We then describe two simple, explicit encoders that can achieve perfect secrecy rates close to these UBs with respect to an untrustworthy relay in the single antenna and single relay setting. Lastly, we generalize our system to a MIMO relay channel where the relay has more antennas than the users and optimal precoding matrices which maintain a required secrecy constraint are studied.  Our results establish that the design of PNC transmission schemes with enhanced throughput and guaranteed data confidentiality is feasible in next generation systems.
\end{abstract}


\IEEEpeerreviewmaketitle

\section{Introduction}
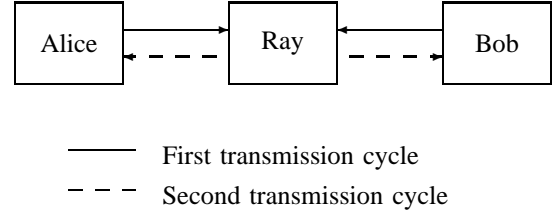
\begin{figure}[t]
\setlength{\unitlength}{0.14in} 
\centering 
\begin{picture}(24,9) 
\put(2,5){\framebox(4,3){Alice}}
\put(10,5){\framebox(4,3){Ray}}
\put(18,5){\framebox(4,3){Bob}}
\put(6,7){\vector(1,0){4}}
\put(18,7){\vector(-1,0){4}}
\put(9, 6){\line(1, 0){0.5}} \put(8, 6){\line(1, 0){0.5}} \put(7, 6){\line(1, 0){0.5}} \put(6.5, 6){\vector(-1, 0){0.5}}
\put(14.5, 6){\line(1, 0){0.5}} \put(15.5, 6){\line(1, 0){0.5}} \put(16.5, 6){\line(1, 0){0.5}} \put(17.5, 6){\vector(1, 0){0.5}}
\put(4, 2.5){\line(1, 0){2.5}}\put(7.5, 2.0){First transmission cycle}
\put(4, 1.0){\line(1, 0){0.5}} \put(5, 1.0){\line(1, 0){0.5}} \put(6, 1.0){\line(1, 0){0.5}}\put(7.5, 0.5){Second transmission cycle}
\end{picture}
\caption{Physical layer network coding (PNC) with two transmitter and one relay node.}
\label{fig:PNC}
\end{figure}

Recently, the ideas of network coding \cite{Ahlswede00} have been extended to the wireless physical medium; notably, in \cite{GastparIT}, \cite{Popovski}, among others, the concept of harnessing interference through structured codes was explored in the framework of physical layer network coding (PNC). These technologies can be proven instrumental in enabling the envisaged multi-fold increase in data throughput in fifth generation ($5$G) networks \cite{Heath14}. The generic PNC system model with two independent sources and one relay is depicted in Fig. \ref{fig:PNC} and assumes that communication is executed in two cycles. In the first cycle, the nodes A, referred to as Alice, and B, referred to as Bob, transmit simultaneously respective codewords to the relay node R, referred to as Ray. In the second cycle, Ray, broadcasts to Alice and Bob a function of the total received signal; Alice and Bob then retrieve each's other messages by canceling off their corresponding transmissions. Depending on the transformation executed by Ray, one of the following relaying strategies can be employed:
amplify and forward, decode and forward, compress and forward \cite{Laneman04}, or the recently introduced compute and forward \cite{Gastpar} approach.

Nevertheless, despite the potential for substantial increase of the transmission rates in wireless networks, a major obstacle in the widespread deployment of PNC and generally of relay networks arises due to security concerns, i.e., the confidentiality of the exchanged data with respect to an untrustworthy relay. A straightforward approach would be employing encryption at upper layers of the communication network or encryption at the physical layer \cite{Chorti10}. However, the management of secret keys used by the crypto algorithms depends on the structure of the access network and already fourth generation systems
($4$G) have a key hierarchy of height five (5) for each individual end-user, while there exist multiple keys in each layer of the hierarchy \cite{WAPstd}. Extrapolating from the experience of $4$G systems, it is expected that the management of secret keys in $5$G would become an even more complicated task \cite{EditorialJSAC13}. The generation, the management and the distribution of secret keys in decentralized settings, such as device-to-device PNC networks, without an infrastructure that supports key management and authentication will impose new security challenges.

An alternative theoretical framework for the study of data confidentiality in the physical layer of wireless networks, dubbed as physical layer security \cite{Bloch_Barros11,PoorBook,ChortiCM}, has recently become a focal point of research in the wireless community. The metric of interest, referred to as the channel secrecy capacity is the supremum of transmission rates at which data can be exchanged reliably while satisfying a weak secrecy \cite{Wyner75}, \cite{Csiszkar78}, a strong secrecy  \cite{Bennett95} or a perfect secrecy constraint \cite{Shannon49}.
 As an example, let $X^n$ be the $n$-length encoded version of a $nR$-bit message
transmitted by the source and let $Z^n$ denote the passive eavesdropper's information. Weak and strong secrecy assume that the code's blocklength $n$ becomes arbitrarily long, while Shannon's definition of perfect secrecy in \cite{Shannon49} on the other hand explicitly assumes a finite blocklength, i.e.,
 \begin{eqnarray}\label{sec_defns}
 &&\lim_{n\rightarrow \infty}\frac{1}{n}I(X^n; Z^n)=0, \text{ weak secrecy constraint},\\
 &&\lim_{n\rightarrow \infty}I(X^n; Z^n)=0, \text{ strong secrecy constraint},\\
 &&I(X; Z)=0, \text{ perfect secrecy constraint}.
 \end{eqnarray}

\begin{figure*}[t]
\setlength{\unitlength}{0.14in} 
\centering 
\begin{picture}(40,11) 
\put(3,7){\framebox(6,3){inner enc. $\theta_A$}}
\put(12,7){\framebox(6,3){outer enc. $\varphi_A$}}
\put(21,7){\framebox(6,3){PNC channel}}
\put(30,7){\framebox(6,3){PNC wrap. $f$}}
\put(28,1){\framebox(8,3){Ray-Bob channel}}
\put(19,1){\framebox(6,3){outer dec. $\phi_B$}}
\put(10,1){\framebox(6,3){inner dec. $\vartheta_B$}}
\put(0,8.5){\vector(1,0){3}}
\put(9,8.5){\vector(1,0){3}}
\put(18,8.5){\vector(1,0){3}}
\put(27,8.5){\vector(1,0){3}}
\put(36,8.5){\vector(1,0){3}}
\put(39,2.5){\line(0,1){6}}
\put(36, 2.5){\line(1, 0){3}}
\put(28, 2.5){\vector(-1, 0){3}}
\put(19, 2.5){\vector(-1, 0){3}}
\put(10, 2.5){\vector(-1, 0){3}}
\put(1,9.5) {$s^q_A$}
\put(10,9.5) {$u_A^n$}
\put(19,9.5){$x_{A}^n$}
\put(28,9.5) {$y^n$}
\put(37, 9.5){$z^n$}
\put(26, 3.5){$y_B^n$}
\put(17, 3.5){$\hat{u}_A^n$}
\put(8, 3.5){$\hat{s}_A^q$}
\end{picture}
\caption{Nested encoder for strong secrecy in PNC systems: Alice encodes the secret messages using an inner encoder $\theta_A$ for reliability and an outer encoder $\varphi_A$ for secrecy. On the other hand, Bob employs the corresponding decoders $\phi_B$ and $\vartheta_B$ to obtain an estimate of the secret messages transmitted by Alice.}
\label{fig:nested}
\end{figure*}
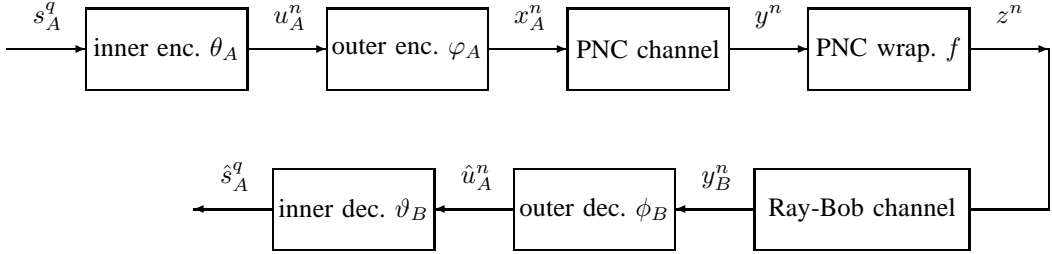

The first study of weak secrecy in relay channels with confidential messages has appeared in \cite{Oohama01} while further analyses followed \cite{Oohama07,He10}; these contributions established that the secrecy capacity of one-way relay channels is zero, unless the source-destination channel is better than the source-relay channel. In essence, relay topologies of practical interest in which the link to the relay is better than the direct link were shown to be inherently insecure. Due to this limiting result, subsequent work focused entirely on cooperative relay channels with trustworthy relays, \cite{Lai08,Lun10,Chen12,Bassily13} to cite but a few.

However, unlike one-way relay networks, systems employing network coding can on the other hand benefit from the simultaneity of transmissions to an untrustworthy relay to achieve data confidentiality as noted in \cite{Vilela08}. In essence, the structured interference observed by the relay can be exploited to achieve strong secrecy in the wireless transmissions \cite{He13,Vatedka15}. In \cite{He08,He09,Ling12}
the role of interference in achieving strong secrecy was demonstrated using lattice encoders; in these works the superposition of the interference to the data was viewed as a modulo addition operation, i.e., the superposition was assumed to take place in the code space and not in the signal space.

In the present study, PNC networks in which Ray can observe superpositions in the signal space (real sums of signals transmitted by Alice and Bob as opposed to modulo sums in the code space) are investigated in the presence of synchronization errors assuming all nodes employ $M$-ary pulse amplitude modulation ($M$-PAM) transceivers; this \textit{realistic} scenario is fundamentally more demanding than previously investigated settings \cite{Bassily13}.
To separate the problem of secrecy from error correction, we first restrict to a noiseless channel where we evaluate upper bounds (UBs) on the achievable perfect secrecy rates, make these explicit in the case of PAM modems, and investigate the effect of synchronization errors on secrecy. 
The proposed secret encoders in the single input single output (SISO) setting are constructive examples of coset coding that allow Ray to obtain estimates of linear combinations of the transmitted PAM symbols but not to retrieve any of the secret bits they carry, thus achieving perfect secrecy, i.e., \emph{zero information leakage per PAM symbol}. Finally, our system model is extended to the multiple input multiple output (MIMO) case in which we study optimal precoding matrices which achieve the required signal alignment at the relay, while preserving secrecy. Our study differs from earlier work on interference alignment for secrecy  \cite{Koyluoglu11}, \cite{Xie12}, \cite{yener2} and interference alignment for the MIMO channel \cite{Bresler14} in that the required secrecy conditions demand equality of \emph{matrices} rather than just of the \emph{subspaces} generated by their columns.

The paper is organized as follows. In Section \ref{sec:system model} the SISO system model is presented.  In Section \ref{sec:SecCapacity} we propose upper bounds (UBs) on the achievable perfect secrecy rates of the noiseless SISO system given finite constellations, provide explicit formulas for these bounds in the case of PAM modems, and further discuss the impact of synchronization errors on the upper bounds.  In Section \ref{sec:enc} two explicit encoders achieving perfect secrecy are constructed, the first assuming no cooperation between the users, and the second assuming that the user of the smaller constellation has some non-trivial information about the signal transmitted by the other user.  Both approaches are shown to be close to the relevant upper bounds.   In Section \ref{sec:MIMO} we generalize our setup to a noisy MIMO channel in which the users and the relay have multiple antennas, and study optimal precoding matrices.  Finally in Section \ref{sec:Conclusions} the conclusions of this contribution are drawn and future directions of the work are discussed.


\section{Secure PNC System Model}\label{sec:system model}
Communication between Alice and Bob with the help of Ray takes place into two cycles as depicted in Fig. \ref{fig:PNC}. In what follows, we use the subscript $A$ to denote quantities and variables (source symbols, codewords, etc.) corresponding to Alice 
 and the subscript $B$ for those belonging to Bob. 
 All channel coefficients and encoding/decoding algorithms are public, i.e., known by Alice, Bob, and Ray. The notation $x^n$ denotes the sequence $[x(1), x(2), \ldots, x(n)]$ while lower case letters denote realizations of respective random variables that are represented with the corresponding upper case letters, e.g., $x$ denotes a realization of the random variable $X$ with probability mass function (pmf) $p_X(x)$.

We assume that Alice's and Bob's source symbols (secret messages) are drawn from discrete alphabets. Under an average power constraint, the use of Gaussian encoders has been demonstrated to achieve the secrecy capacity of the interference channel \cite{PoorInterferer}. However when transmission is constrained by a joint amplitude-variance constraint\footnote{Under this realistic assumption the amplitude of the transmitted signals is bounded, as in all actual communication systems.} it has been shown that the capacity is on the contrary achieved by employing codebooks of finite size; a recent extension of these results in the wiretap channel has shown that this holds true for the secrecy capacity as well \cite{Ozel14}. Due to this reason, in the following we exclusively operate under the assumption that all codebooks have finite size.

We start by examining the scenario in which all nodes have single antennas while the multi-antenna case will be covered in a later section.
 In the present work we treat separately the design of secrecy encoders from error correction encoding; this can be straightforwardly achieved with the nested structure depicted in Fig. \ref{fig:nested} including an inner encoder for reliability and an outer encoder for secrecy. The reason we propose this approach is that, contrary to error correction, in the noiseless PNC setting it is possible to achieve perfect secrecy without introducing \textit{any delay}, i.e., secrecy is achieved on a per symbol basis and does not rely on the existence of noise to increase the equivocation at Ray but rather on structured interference.

In the presentation of the proposed nested encoder we employ the following notation: Alice (respectively Bob) employs a rate $\frac{q}{n}, q, n\in \mathds{N}, q\leq n$ \textit{inner encoder} for reliability denoted by $\theta_A$ (respectively $\theta_B$) and corresponding decoder denoted by $\vartheta_B$ (respectively $\vartheta_A$). Furthermore, to ensure perfect secrecy Alice (Bob) uses a unit symbol rate \textit{outer secrecy encoder} denoted by $\varphi_A$ (respectively $\varphi_B$) with a corresponding decoder $\phi_B$ (respectively $\phi_A$).
On the other hand Ray employs a PNC wrapping function (e.g., ``compress and forward'') denoted by $f$.
In subsections \ref{subsec: first cycle} to \ref{subsec:perfect secrecy} we explain the above setting in further detail and discuss the necessary secrecy conditions.

\subsection{First transmission cycle}\label{subsec: first cycle}
\textit{Inner encoder for reliability}: In the first cycle Alice maps length-$q$ sequences of secret messages $s_A\in \mathcal{S}_A$, selected uniformly at random from the set $\mathcal{S}_A$ of secret messages to length-$n$ codewords. To this end, Alice first employs an encoder $\theta_A: \mathcal{S}_A^q \rightarrow \mathcal{U}_A^n$, $\theta_A(s_A^q) = u_A^n$. Similarly Bob maps length-$q$ sequences of secret messages $s_B\in \mathcal{S}_B$ using an encoder $\theta_B: \mathcal{S}_B^q \rightarrow \mathcal{U}_B^n$, $\theta_B(s_B^q) = u_B^n$.

\textit{Outer encoder for perfect secrecy}: Alice and Bob encode the sequences ${u}_A^n$ and ${u}_B^n$ respectively to codewords $x_A^n$
  and $x_B^n$ element by element using corresponding encoders $\varphi_A: \mathcal{U}_A\rightarrow \mathcal{X}_A$ and $\varphi_B: \mathcal{U}_B\rightarrow
  \mathcal{X}_B$ with $\varphi_A(u_A)=x_A$ and $\varphi_B(u_B)=x_B$. We define $M_A$ and $M_B$ to be the sizes of the codebooks $\mathcal{X}_A$ and $\mathcal{X}_B$, denoted by $|\mathcal{X}_A| = M_A$ and $|\mathcal{X}_B| = M_B$.  We set $m_A = \log_2M_A$ and $m_B = \log_2M_B$.

Ray's observation (assuming perfect synchronization at Alice and Bob) can be expressed as
\begin{eqnarray}
y = h_A x_A + h_B x_B + w_R = x_R + w_R,
\end{eqnarray}
where $h_A$ (respectively $h_B$) is the channel (fading) coefficient in the link from Alice (Bob) to Ray, $x_R = h_Ax_A+ h_Bx_B$, and $w_R$ is the noise at Ray, modeled as (a realization of) a zero-mean circularly symmetric Gaussian complex random variable with variance $\sigma_R^2$.

\subsection{Second transmission cycle} \label{subsec: second cycle}
\textit{PNC wrapping}: In the second cycle of the communication, Ray performs a wrapping of the received PNC observation $y$ to compressed PNC observations using a mapping $f:\mathcal{Y}\rightarrow\mathcal{Z}$, where $f({y}) = z$ (e.g., possible options for this mapping include ``compress and forward'' and ``compute and forward''). We assume that $f$ is invertible given either $x_A$ or $x_B$, that is, that Alice can recover $y$ from $z$ given that she knows $x_A$, and similarly for Bob.  An obvious choice is to select $\mathcal{Z} = \mathcal{Y}$ and have $f$ be the identity function, i.e., Ray forwards exactly what he receives. 

Finally, Ray transmits $z$ to Alice and Bob, who then observe
\begin{eqnarray}
y_A= \tilde{h}_Az+w_A, \label{output_alice}\\
y_B= \tilde{h}_Bz+w_B, \label{output_bob}
\end{eqnarray}
where $\tilde{h}_A$ (respectively $\tilde{h}_B$) is the channel (fading) coefficient from Ray to Alice (Ray to Bob), and $w_A$ ($w_B$) is the noise at Alice (Bob), modeled as a circularly symmetric complex Gaussian random variable with variance $\sigma_A^2$ ($\sigma_B^2$).

Alice uses a decoder $\phi_A:  \mathcal{Y}_A \rightarrow \mathcal{U}_B$, to produce estimates $\phi_A(y_A) = \hat{u}_B$ of Bob's transmitted secret codewords. Respectively, Bob uses a function $\phi_B:  \mathcal{Y}_B \rightarrow \mathcal{U}_A$, to produce estimates $\phi_B(y_B) = \hat{u}_A$ of Alice's transmitted secret codewords.

For the purposes of error correction Alice (Bob) employs a decoding function $\vartheta_A: \mathcal{U}_B^n\rightarrow \mathcal{S}_B^q$ with $\vartheta_A(\hat{u}_B^n)=\hat{s}_B^q$ ($\vartheta_B: \mathcal{U}_A^n\rightarrow \mathcal{S}_A^q$ with $\vartheta_B(\hat{u}_A^n)=\hat{s}_A^q$). Focusing exclusively on the secrecy of the PNC scheme, we assume that $\lim_{n\rightarrow \infty}P_{e,A}^n=0$ ($\lim_{n\rightarrow \infty}P_{e,B}^n=0$) where the probability of a decoding error at Alice (respectively Bob) is $P_{e,A}^n=\mathrm{Pr}[\vartheta_A(\hat{u}_B^n)\neq s_B^q]$ (respectively ($P_{e,B}^n=\mathrm{Pr}[\vartheta_B(\hat{u}_A^n)\neq s_A^q]$).

\subsection{Perfect Secrecy and an Upper Bound on the Achievable Secrecy Rates}\label{subsec:perfect secrecy}

Perfect secrecy can be achieved with respect to Ray if the mutual information between Ray's observation and the secret source symbols is zero, i.e.,
\begin{eqnarray}
I(Y; S_A) &=&0, \text{ perfect secrecy condition for Alice} \label{eq:perfect secrecy Alice}\\
 I(Y; S_B) &=&0, \text{ perfect secrecy condition for Bob}\label{eq:perfect secrecy Bob}.
\end{eqnarray}
The input and output random variables in the PNC system model form respective Markov chains $S_A^q\rightarrow U_A^n\rightarrow X_A^n \rightarrow X_R^n\rightarrow Y^n$ and $S_B^q\rightarrow U_B^n\rightarrow X_B^n \rightarrow X_R^n\rightarrow Y^n$. As a result, due to the data processing inequality, to satisfy conditions (\ref{eq:perfect secrecy Alice}) and (\ref{eq:perfect secrecy Bob}) it suffices to show that
\begin{eqnarray}
I(X_R; S_A)&=&0, \text{ sufficient condition for (\ref{eq:perfect secrecy Alice}),}\label{eq:equivalent perfect secrecy Alice}\\
I(X_R; S_B)&=&0, \text{ sufficient condition for (\ref{eq:perfect secrecy Bob})} \label{eq:equivalent perfect secrecy Bob}.
\end{eqnarray}

For fixed input distributions, encoders, and channels, we will study the perfect secrecy rates\footnote{As the messages are delivered over two transmission cycles, one should potentially multiply all upper bounds and rates by $\frac{1}{2}$; however we omit this factor for clarity and as it does not affect the nature of our results.} $R^s_A$ and $R^s_B$ provided (\ref{eq:equivalent perfect secrecy Alice}) and (\ref{eq:equivalent perfect secrecy Bob}) are satisfied.  To measure the optimality of our encoding schemes, we will compare them with the following upper bounds (UBs) on the achievable perfect secrecy rates:
\begin{align}
R^s_A \leq \widehat{R}^s_A &= \left[I(Y_B; X_A|X_B) - I(Y; X_A)\right]^+, \label{eq:SC Alice}\\
R^s_B \leq \widehat{R}^s_B &= \left[I(Y_A; X_B|X_A) - I(Y; X_B)\right]^+. \label{eq:SC Bob}
\end{align}
In the following section we investigate $\widehat{R}^s_A$ and $\widehat{R}^s_B$ further for noiseless channels.  In \cite{Csiszkar78} it was only shown that (\ref{eq:SC Alice}) and (\ref{eq:SC Bob}) are achievable for weak secrecy as in (\ref{sec_defns}).  As we consider only perfect secrecy, the above serve only as upper bounds.  However, their intuitive and easily-computable nature yields them useful nonetheless.


\section{Upper Bounds in the Noiseless Scenario} \label{sec:SecCapacity}
Throughout this section and the next we assume that 
\begin{itemize}
\item[(i)] the pmfs of $X_A$ and $X_B$ are uniform, and
\item[(ii)] all channels are fixed and invertible.
\end{itemize}

When channel state information is globally available, we assume that Alice and Bob employ channel precoders, denoted respectively by $g_A$ and $g_B$, such that
 \begin{equation}
h_Ag_A = h_Bg_B
 \end{equation}
 so that Ray observes
 \begin{equation}\label{ray_observe}
 y =h_Ag_Ax_A+h_Bg_Bx_B + w_R =h_Ag_A(x_A+x_B) + w_R.
 \end{equation}
Ray now attempts to recover the sum $x_A + x_B$.  The secrecy of our proposed encoders depends only on the structure of the sum $x_A+x_B$, so we set $w_R = 0$ in the next two sections.  In this noiseless environment, Ray can post-multiply the received signal in (\ref{ray_observe}) by $(h_Ag_A)^{-1}$ to recover $x_A+x_B$ exactly.  Similarly, the Ray-Alice and Ray-Bob channels are assumed noiseless.  We summarize by adding a third assumption:
\begin{itemize}
\item[(iii)] All channel gains are equal to unity and all noise sources are zero, that is, $h_A = h_B = \tilde{h}_A = \tilde{h}_B = 1$ and $w_A = w_B = w_R = 0$.
\end{itemize}
While this may seem unrealistic, we are rather interested in the achievable perfect secrecy rates based solely on the structure of the sum $x_A+x_B$ itself.  Thus while the presence of noise and channel gains can have a deteriorating effect on Alice and Bob's overall data rate, removing assumption (iii) will not affect perfect secrecy relative to Ray, nor will it affect the perfect secrecy rate relative to the overall data rate.
 
We note that although channel inversion is impractical in Rayleigh environments, it can be employed whenever a line of sight (LOS) exists between either transmitter and Ray, i.e., whenever a Rician, a Nakagami-m or other large scale fading channel model \cite{Hashemi} is applicable.  We will return to the question of designing optimal precoders $g_A$ and $g_B$ in the presence of noise in Section \ref{sec:MIMO}.

\subsection{An Upper Bound on the Achievable Perfect Secrecy Rate}\label{sec:noiseless}

In the noiseless setting with unit channel gains, the set of all possible observations at Ray is
\begin{equation}
\mathcal{Y} = \mathcal{X}_R = \{x_A + x_B\ |\ x_A \in\mathcal{X}_A,x_B\in\mathcal{X}_B\}
\end{equation}
which comes with an addition function
\begin{equation}
\psi :\mathcal{X}_A \times \mathcal{X}_B\rightarrow \mathcal{Y},\quad \psi(x_A,x_B) = x_A + x_B.
\end{equation}
Crucial to our analysis are the sets
\begin{equation}
\psi^{-1}(y) = \{(x_A,x_B)\ |\ y = x_A + x_B\}.
\end{equation}
The pmf of $Y$ is given by the convolution of the pmfs of $X_A$ and $X_B$, which is clearly seen to be
\begin{equation}\label{ray_pmf}
p_Y(y) = \sum_{\substack{ x_A, x_B \\ x_A + x_B = y}}p_{X_A}(x_A)p_{X_B}(x_B) = \frac{|\psi^{-1}(y)|}{M_AM_B}.
\end{equation}
The following proposition gives a compact, intuitive upper bound on the achievable secrecy rates $\widehat{R}^s_A$ and $\widehat{R}^s_B$.
\begin{prop}\label{sec_cap_explicit}
In the noiseless scenario $\widehat{R}^s_A$ and $\widehat{R}^s_B$ are equal.   Furthermore, denoting $\widehat{R}^s = \widehat{R}^s_A = \widehat{R}^s_B$, we have
\begin{equation}
\widehat{R}^s = \sum_{y\in\mathcal{Y}}\log_2|\psi^{-1}(y)|\frac{|\psi^{-1}(y)|}{M_AM_B}
\end{equation}
\end{prop}
\begin{IEEEproof}
See Appendix \ref{app: Proof of Proposition 1}.
\end{IEEEproof}

In the noiseless scenario we therefore simply define $\widehat{R}^s$ to be \emph{the} perfect secrecy rate UB.  Intuitively, for a given $y\in \mathcal{Y}$, $\log_2|\psi^{-1}(y)|$ measures equivocation at Ray in bits, and should therefore upper bound the total number of secret bits that Alice and Bob can transmit when Ray observes $y$ while maintaining perfect secrecy.  The quantity $|\psi^{-1}(y)|/M_AM_B = p_Y(y)$ measures the frequency at which Ray observes $y$, properly weighting the sum as a rate calculation would.  However, note that to fully exploit the value of $|\psi^{-1}(y)|$, Alice and Bob would need to know this number, and hence need non-trivial knowledge of each other's symbols, in advance.  


\subsection{Guaranteed Entropy}\label{sec:gent}

In this subsection, we establish upper bounds on the achievable secrecy rates in the scenario in which Alice and Bob have no knowledge of the other's symbols.  For such a scenario, these upper bounds are necessarily tighter than the above $\widehat{R}^s$.   First, we define for any symbols $x_A$ and $x_B$ a useful notion of the amount of confusion Ray is guaranteed to experience when transmitting one of these points.

\begin{definition}
For any $x_A \in \mathcal{X}_A$ we define the \emph{guaranteed entropy} of $x_A$ to be
\begin{equation}
s(x_A) = \min_{x_B\in \mathcal{X}_B} \log_2|\psi^{-1}(x_A + x_B)|.
\end{equation}
and similarly for any $x_B\in \mathcal{X}_B$ we define
\begin{equation}
s(x_B) = \min_{x_A\in \mathcal{X}_A} \log_2|\psi^{-1}(x_A + x_B)|.
\end{equation}
\end{definition}

The following proposition is more or less immediate.

\begin{prop}\label{upper_bound2}
In the noiseless scenario with unit channel gains where Alice has no knowledge of $x_B$ and Bob has no knowledge of $x_A$, the achievable secrecy rates are upper bounded by $R^s_A\leq \widetilde{R}^s_A$ and $R^s_B\leq \widetilde{R}^s_B$ where
\begin{equation}
\widetilde{R}^s_A = \sum_{x_A\in \mathcal{X}_A}\frac{s(x_A)}{M_A},\ \  \widetilde{R}^s_B = \sum_{x_B\in \mathcal{X}_B}\frac{s(x_B)}{M_B}
\end{equation}
\end{prop}
\begin{IEEEproof}
Suppose that a secret message $s_A$ of length $l(s_A)$ is encoded in $x_A$, so that $\log_2l(s_A)$ measures the amount of information in $s_A$ in bits.  For any $x_B$, Ray can determine $m_A-\log_2|\psi^{-1}(x_A+x_B)|$ of the total number $m_A$ of bits transmitted by Alice.  As Alice has no knowledge of $x_B$, the value of $l(x_A)$ must be independent of $x_B$, and it follows that maintaining perfect secrecy requires $\log_2l(s_A)\leq \log_2|\psi^{-1}(x_A+x_B)|$ for all $x_B$.  Hence $\log_2l(x_A)\leq s(x_A)$, from which it follows that
\begin{equation}
R^s_A \leq \sum_{x_A\in \mathcal{X}_A}\log_2l(s_A)\cdot p_{X_A}(x_A) \leq \sum_{x_A\in \mathcal{X}_A}\frac{s(x_A)}{M_A}
\end{equation}
as claimed.  Identical reasoning applies to Bob.
\end{IEEEproof}

Note that the above upper bounds $\widetilde{R}^s_A$ and $\widetilde{R}^s_B$ hold regardless of the nature of the input distributions $S_A$ and $S_B$; that is, they apply to non-binary and non-uniform secret input distributions alike.  Similarly, they hold regardless of whether one chooses to code over several time instances.


\subsection{Upper Bounds on the Achievable Perfect Secrecy Rates for PAM Modems}\label{sec:pam_achieve}
Let us now study a familiar scenario in which $\mathcal{X}_A$ and $\mathcal{X}_B$ are, respectively, $M_A$- and $M_B$-PAM constellations, so that $X_A$ is the uniform distribution on
\begin{equation}
\mathcal{X}_A = \{-(M_A-1),-(M_A-3),\ldots,M_A-3,M_A-1\}
\end{equation}  
and similarly for $X_B$.  Throughout this section we assume that $M_B\geq 2M_A$.  The theorems of this subsection evaluate the upper bounds of the previous subsection for PAM modems.

\begin{prop}\label{formula_for_psiy}
In the noiseless setting with unit channel gains when Alice and Bob employ $M_A$-PAM and $M_B$-PAM modulators with $M_B\geq 2M_A$, we have
\begin{align}
&|\psi^{-1}(y)| = \nonumber\\
& \left\{
\begin{array}{ll}
0 & \text{$y$ odd or $|y|\geq M_B+M_A$}, \\
\frac{M_B + M_A - |y|}{2}  & M_B - M_A +2 \leq |y| \leq M_B + M_A - 2, \\
M_A & |y|   \leq M_B - M_A. \\
\end{array}
\right.\nonumber
\end{align}
\end{prop}
\begin{IEEEproof}
This is a straightforward calculation and is therefore omitted.
\end{IEEEproof}

\begin{figure}[t]
\centering
\includegraphics[angle=0,width = 0.4\textwidth]{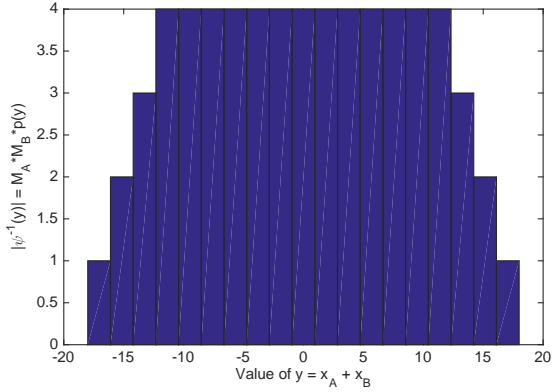}
\caption{The pmf of Ray's observation $y = x_A + x_B$ in the noiseless scenario with channel gains $h_A = h_B = 1$.  Here Alice employs a $4$-PAM modulator and Bob a $16$-PAM modulator.} \label{fig:rays_pmf_pam}
\end{figure}

The above proposition in conjunction with (\ref{ray_pmf}) allows us to explicitly describe the Ray's pmf $p_Y(y)$.  In Fig.\ \ref{fig:rays_pmf_pam} we plot the values of $|\psi^{-1}(y)|$ for $M_A = 4$ and $M_B = 16$.  In the following sections we will exploit the  trapezoidal nature of Ray's pmf when constructing explicit encoding functions.

\begin{theorem}\label{pam_cap}
In the noiseless setting with unit channel gains when Alice and Bob employ $M_A$-PAM and $M_B$-PAM modulators with $M_B\geq 2M_A$, we have 
\begin{equation}
\widehat{R}^s = m_A\frac{M_B-M_A+1}{M_B} + \sum_{a = 1}^{M_A-1}\log_2(a)\frac{2a}{M_AM_B}. \label{eq:UB}
\end{equation}
In particular, for fixed $M_A$ we have $\lim_{M_B\rightarrow\infty}\widehat{R}^s = m_A$.
\end{theorem}
\begin{IEEEproof}
See Appendix \ref{app: Proof of Theorem 1}.
\end{IEEEproof}

The follow proposition explicitly computes $s(x_A)$ and $s(x_B)$, which immediately results in explicit expressions for $\widetilde{R}^s_A$ and $\widetilde{R}^s_B$ of Section \ref{sec:gent} which do not assume cooperation between Alice and Bob.  

\begin{prop}\label{explicit_gent}
In the noiseless setting with unit channel gains when Alice and Bob employ $M_A$-PAM and $M_B$-PAM modulators with $M_B\geq 2M_A$, the guaranteed entropies of $x_A$ and $x_B$ are given by
\begin{align}
s(x_A) &= \log_2\left(\frac{M_A+1-|x_A|}{2}\right) \\
s(x_B) &= \left\{\begin{array}{cl}
m_A & |x_B|\leq M_B-2M_A+1 \\
\log_2\left(\frac{M_B+1-|x_B|}{2}\right) & \text{otherwise.} \nonumber
\end{array}\right.
\end{align}
\end{prop}
\begin{IEEEproof}
See Appendix \ref{app:proof of prop 4}.
\end{IEEEproof}




\subsection{Effect of Time Synchronization Errors on the Upper Bounds}
\begin{figure}[t]
\centering
\includegraphics[angle=0,width=0.50\textwidth]{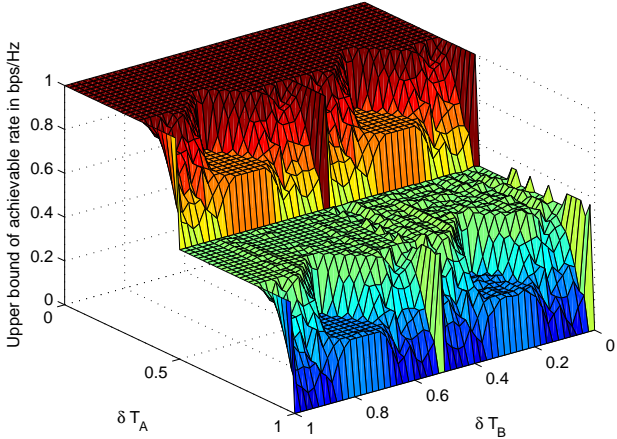}
\caption{Numerical evaluation of the UB $\widehat{R}_s$ in the presence of time synchronization errors denoted by $\delta T_A$ and $\delta T_B$. The symbol period is normalized to unity $T=1$.}\label{fig: sync_err}
\end{figure}

One of the main issues in PNC networks is that the assumption of perfect time synchronization is too optimistic. In this subsection we investigate the effect of time synchronization on $\widehat{R}_s$ when Alice and Bob employ $M$-PAM modulators. In this case the analog signals transmitted by Alice and Bob, denoted by $t_A(t), \text{ and }t_B(t)$ respectively, can be expressed as:
\begin{eqnarray}
t_A(t)&=&\sum_{l=-\infty}^{\infty}{x_A(l) \cos(2\pi ft )g(t-lT-\delta T_A)},\\
t_B(t)&=&\sum_{l=-\infty}^{\infty}{x_B(l) \cos(2\pi ft )g(t-lT-\delta T_B)},
\end{eqnarray}
where $T$ is the symbol period, $\delta T_A$ and $\delta T_B$ denote the synchronization errors at Alice and Bob respectively and are assumed to be uniformly distributed in the range $[0, T]$ and $g(t)$ denotes the transmitter filter (commonly implemented as a raised cosine filter). Here for simplicity we assume that the transmitter filter is a simple rectangular window of length equal to the symbol period $T$.
Neglecting all other noise sources and assuming that Ray employs a matched filter receiver implemented as a standard correlator receiver, Ray's observation $y(l)$ during the $l$-th symbol can be expressed as
\begin{eqnarray}
y(l)&=&\frac{2}{T}\int_{-\delta T_A}^{0}{x_A(l-1)\cos^2 (2 \pi ft) dt}\nonumber\\
&+& \frac{2}{T}\int_{0}^{T-\delta T_A}{x_A(l)\cos^2 (2 \pi ft) dt}\nonumber\\
&+& \frac{2}{T}\int_{-\delta T_B}^{0}{x_B(l-1)\cos^2 (2 \pi ft) dt}\nonumber\\
&+& \frac{2}{T}\int_{0}^{T-\delta T_B}{x_B(l)\cos^2 (2 \pi ft) dt}\nonumber\\
&=&(1-\alpha)x_A+(1-\beta)x_B+\alpha x_A(l-1) \nonumber\\
&+&\beta x_B(l-1),
\end{eqnarray}
where
\begin{eqnarray}
\alpha&=&\frac{\sin\left(4 \pi \frac{\delta T_A }{T}\right)}{4 \pi}+\frac{\delta T_A }{T},\\
\beta&=&\frac{\sin\left(4 \pi \frac{\delta T_B }{T}\right)}{4 \pi}+\frac{\delta T_B }{T}.
\end{eqnarray}
As a result of time synchronization errors, Alice's and Bob's symbols are misaligned when reaching Ray. We investigate the effect of this misalignment on the UBs by numerically evaluating (\ref{eq:SC Alice}) and (\ref{eq:SC Bob}). The results are depicted results in Fig. \ref{fig: sync_err}; for relatively small synchronization errors $\delta T_A, \delta T_B<0.25T$, the effect on $\widehat{R}_s$ is negligible. On the other hand, as the synchronization errors increase their impact on the UBs becomes increasingly important. Interestingly, due to the sinusoidal parts of $\alpha$ and $\beta$, there are four regions of values of $(\delta T_A, \delta T_B)$ -- around the points $\left(\frac{1}{4}, \frac{1}{4}\right), \left(\frac{1}{4}, \frac{3}{4}\right), \left(\frac{3}{4}, \frac{1}{4}\right) \text{ and } \left(\frac{3}{4}, \frac{3}{4}\right)$ -- in which the decrease in $\widehat{R}^s$ is more acute.

\section{Explicit Encoder Construction with PAM Modems in the Noiseless Scenario}\label{sec:enc}

Throughout this section we assume that Alice and Bob use PAM modems with sizes $M_A$ and $M_B$, respectively, satisfying $M_B\geq 2M_A$.  We retain the assumptions of the previous section, namely that all channels are noiseless and all channel gains are set to unity.  


The central idea behind the proposed approach in designing the secrecy encoders $\varphi_A$ and $\varphi_B$ stems from the following observation. The superposition of two PAM signals is equivalent to the convolution of two uniform pmfs; the resulting pmf contains a ``flat" region in which Ray's observations are equiprobable and two ``linear" regions in which combinations of symbols occur with increasing/decreasing probabilities, as demonstrated in Fig. \ref{fig:rays_pmf_pam} for $M_A = 4$ and $M_B = 16$.

In this section we construct explicit secret bit encoders at both Bob and Alice, compute the corresponding perfect secrecy rates, and compare the results to the relevant upper bounds of the previous section.  The first subsection is devoted to the situation wherein neither Alice nor Bob has any information about the other's transmitted signal; thus comparison of the obtained secrecy rates with the upper bounds $\widetilde{R}^s_A$ and $\widetilde{R}^s_B$ is appropriate.  The scheme is shown to perform close to these upper bounds, but Alice's rate is somewhat deficient in an absolute sense, in that as $M_A\rightarrow\infty$, it leaves a constant gap to the optimal asymptotic behavior of $m_A$ bits/sec/Hz.

To increase Alice's secrecy rate, in the second subsection we study her achievable secrecy rate when she has knowledge of $\lfloor s(x_B) \rfloor$ (where $\lfloor \cdot \rfloor$ denotes the floor function) prior to the transmission of this symbol by Bob.  We construct a simple scheme which rectifies the deficient asymptotic performance of the scheme of the previous subsections.

\subsection{Explicit Encoder Construction}

This subsection presents an encoding scheme for both Alice and Bob which does not depend on either user having any knowledge of the other's symbols, i.e.\ no cooperation is necessary between Alice and Bob.  For simplicity we only describe Bob's encoding function; obvious remarks apply to Alice throughout.  We construct our encoding function by first describing a bit labeling procedure on each of Bob's constellation points.  We then declare certain bits of $x_B$ to be secret and the rest to be public, depending essentially on the value of $s(x_B)$, which in turn effectively determines the encoding procedure.

We begin by defining subsets $\mathcal{X}_A^{(k)}$ and $\mathcal{X}_B^{(k)}$ of Alice and Bob's constellations, respectively, for each index $k = 0,\ldots,m_A$, whose usefulness is made clear by the subsequent proposition.  Recall that $M_B\geq 2M_A$.

For Alice's constellation, we set
\begin{equation}
\begin{aligned}
\mathcal{X}_A^{(0)} &= \{x_A\ |\ M_A-5 < |x_A| \leq M_A-1\} \nonumber \\
\mathcal{X}_A^{(k)} &= \{x_A\ |\ M_A-1-2^{k+2}<|x_A|\leq M_A-1-2^{k+1}\} \nonumber \\
&\quad\ \text{for $k = 1,\ldots,m_A-2$, and }\nonumber \\
\mathcal{X}_A^{(k)} &= \emptyset \text{ for $k = m_A-1,m_A$} \nonumber
\end{aligned}
\end{equation}
For Bob's constellation, we define
\begin{equation}
\begin{aligned}
\mathcal{X}_B^{(0)} &= \{x_B\ |\ M_B-5 < |x_B| \leq M_B-1\} \nonumber \\
\mathcal{X}_B^{(k)} &= \{x_B\ |\ M_B-1-2^{k+2}<|x_B|\leq M_B-1-2^{k+1}\}\nonumber \\
&\quad \ \text{for $k = 1,\ldots,m_A-1$, and }\nonumber \\
\mathcal{X}_B^{(m_A)} &= \{x_B\ |\ |x_B|\leq M_B-2M_A-1\} \nonumber
\end{aligned}
\end{equation}

\begin{prop}\label{gent}
The subsets $\mathcal{X}_A^{(k)}$ and $\mathcal{X}_B^{(k)}$ satisfy
\begin{align}
s(x_A) &\geq k\quad \text{for all $x_A\in \mathcal{X}_A^{(k)}$} \\
s(x_B) &\geq k\quad \text{for all $x_B\in \mathcal{X}_B^{(k)}$}
\end{align}
and their cardinalities are given by
\begin{equation}
|\mathcal{X}_A^{(k)}| = 2^{k+1}\quad \text{if $1\leq k\leq m_A-2$}
\end{equation}
and
\begin{equation}
|\mathcal{X}_B^{(k)}| = \left\{
\begin{array}{cl}
2^{k+1} & \text{if $1\leq k \leq m_A-1$} \\
M_B - 2M_A & \text{if $k = m_A$}
\end{array}
\right.
\end{equation}
\end{prop}
\begin{IEEEproof}
This is a straightforward but lengthy application of the formulas for $s(x_A)$ and $s(x_B)$ given in Proposition \ref{explicit_gent}.\end{IEEEproof}

The set $\mathcal{X}_B^{(m_A)}$ is maximal in the sense that one can easily show $s(x_B)< m_A$ for all $x_B\not\in \mathcal{X}_B\cup \{\pm(M_B-2M_A+1)\}$.  Thus $\mathcal{X}_B^{(m_A)}$ is the largest subset of $\mathcal{X}_B$ on which Ray is guaranteed to experience the maximum number of bits of equivocation, namely $m_A$, and whose cardinality is divisible by $M_A$.  This latter condition is necessary to maintain the assumed uniformity of the random variables $S_B$ and $X_B$.  Similar remarks apply for the other $\mathcal{X}_B^{(k)}$.

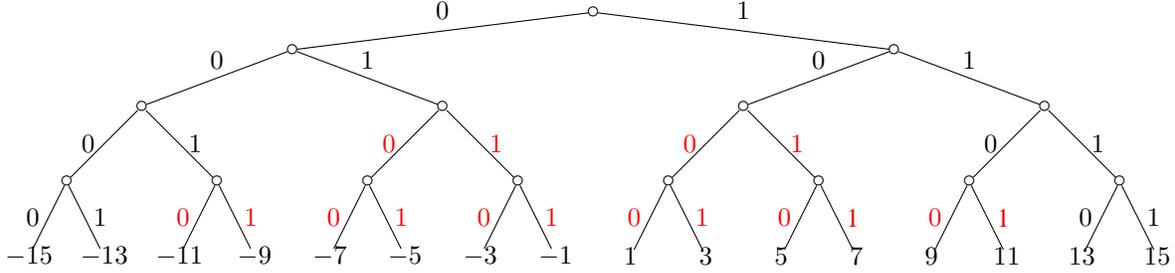
\begin{figure*}[width=\textwidth]
\centering
\begin{tikzpicture}
 \node [outer sep = -4pt]{$\circ$} [level distance=5mm,sibling distance=80mm]
child {
	node [outer sep = -4pt]{$\circ$} [level distance=7.5mm ,sibling distance=40mm]
	child {node [outer sep = -4pt]{$\circ$} [level distance = 10mm, sibling distance = 20mm]
		child {node [outer sep = -4pt] {$\circ$} [level distance = 10mm, sibling distance = 10mm]
			child {node [outer sep = -4pt] {$-15$} edge from parent node [left]{$0$}}
			child {node [outer sep = -4pt] {$-13$} edge from parent node [right]{$1$}}
			edge from parent node [left]{$0$}
			}
		child {node [outer sep = -4pt] {$\circ$} [level distance = 10mm, sibling distance = 10mm]
			child {node [outer sep = -4pt] {$-11$} edge from parent node [left]{$\textcolor{red}{0}$}}
			child {node [outer sep = -4pt] {$-9$} edge from parent node [right]{$\textcolor{red}{1}$}}
			edge from parent node [right]{$1$}
			}
		edge from parent node [left = .01em, above = .01em]{$0$}
		}
	child {node [outer sep = -4pt]{$\circ$} [level distance = 10mm, sibling distance = 20mm]
		child {node [outer sep = -4pt] {$\circ$} [level distance = 10mm, sibling distance = 10mm]
			child {node [outer sep = -4pt] {$-7$} edge from parent node [left]{$\textcolor{red}{0}$}}
			child {node [outer sep = -4pt] {$-5$} edge from parent node [right]{$\textcolor{red}{1}$}}
			edge from parent node [left]{$\textcolor{red}{0}$}
			}
		child {node [outer sep = -4pt] {$\circ$} [level distance = 10mm, sibling distance = 10mm]
			child {node [outer sep = -4pt] {$-3$} edge from parent node [left]{$\textcolor{red}{0}$}}
			child {node [outer sep = -4pt] {$-1$} edge from parent node [right]{$\textcolor{red}{1}$}}
			edge from parent node [right]{$\textcolor{red}{1}$}
			}
		edge from parent node [= .01em,above = .01em]{$1$}
		}
	edge from parent node [above = .1em]{$0$}
	}
child {
	node [outer sep = -4pt]{$\circ$} [level distance=7.5mm ,sibling distance=40mm]
	child {node [outer sep = -4pt]{$\circ$} [level distance = 10mm, sibling distance = 20mm]
		child {node [outer sep = -4pt] {$\circ$} [level distance = 10mm, sibling distance = 10mm]
			child {node [outer sep = -4pt] {$1$} edge from parent node [left]{$\textcolor{red}{0}$}}
			child {node [outer sep = -4pt] {$3$} edge from parent node [right]{$\textcolor{red}{1}$}}
			edge from parent node [left]{$\textcolor{red}{0}$}
			}
		child {node [outer sep = -4pt] {$\circ$} [level distance = 10mm, sibling distance = 10mm]
			child {node [outer sep = -4pt] {$5$} edge from parent node [left]{$\textcolor{red}{0}$}}
			child {node [outer sep = -4pt] {$7$} edge from parent node [right]{$$\textcolor{red}{1}$$}}
			edge from parent node [right]{$\textcolor{red}{1}$}
			}
		edge from parent node [left = .01em, above = .01em]{$0$}
		}
	child {node [outer sep = -4pt]{$\circ$} [level distance = 10mm, sibling distance = 20mm]
		child {node [outer sep = -4pt] {$\circ$} [level distance = 10mm, sibling distance = 10mm]
			child {node [outer sep = -4pt] {$9$} edge from parent node [left]{$\textcolor{red}{0}$}}
			child {node [outer sep = -4pt] {$11$} edge from parent node [right]{$$\textcolor{red}{1}$$}}
			edge from parent node [left]{$0$}
			}
		child {node [outer sep = -4pt] {$\circ$} [level distance = 10mm, sibling distance = 10mm]
			child {node [outer sep = -4pt] {$13$} edge from parent node [left]{$0$}}
			child {node [outer sep = -4pt] {$15$} edge from parent node [right]{$1$}}
			edge from parent node [right]{$1$}
			}
		edge from parent node [right = .01em,above = .01em]{$1$}
		}
	edge from parent node [above = .1em]{$1$}
	};
\end{tikzpicture}
\caption{Secret bit encoder at Bob for $M_A = 4$ and $M_B = 16$, where secret bits are denoted in red and public bits in black.  The subsets $\mathcal{X}_B^{(k)}$ are given by $\mathcal{X}_B^{(2)} = \{-7,\ldots,7\}$ and $\mathcal{X}_B^{(1)} = \{-11,-9,9,11\}$.}\label{tree_encoder}
\end{figure*}

Our bit labeling procedure can be described pictorially by a perfect binary tree as in Fig.\ \ref{tree_encoder}.  Each point in $\mathcal{X}_B$ is assigned in increasing order to a leaf in the perfect binary tree with $M_B$ leaves.  The edges at each level of the tree are alternately labeled with a $0$ or a $1$.  A point $x_B$ is then given a bit labeling by tracing the tree downwards from the root node to the corresponding leaf, so that the bit closest to the root node is the left-most bit in the string.  For the example in Fig.\ \ref{tree_encoder} with $M_B = 16$, we have the bit labelings $0101 \rightarrow -5$, $1011\rightarrow 7$, etc.  It is easy to prove by induction on $m_B$ that each point is assigned a unique bit string of length $m_B$.  

We now declare the last $k$ bits of all $x_B\in \mathcal{X}_B^{(k)}$ to be secret.  To encode queues of public and secret bits, Bob begins at the root node of the tree and travels downwards, encoding public bits until he hits a node all of whose descending edges correspond to secret bits.  He then switches to his queue of secret bits and begins encoding those, until the constellation point to be sent is completely determined.


\begin{example}
Let us set $M_A = 4$ and $M_B = 16$, and suppose that Bob wishes to transmit the public and secret (respectively) bit strings 
\begin{equation}
\mathcal{P}_B = 00110110,\quad \mathcal{S}_B = \textcolor{red}{1001}
\end{equation}
He begins by encoding the left-most bits in $\mathcal{P}_B$, namely $001$, at which point the first bit in $\mathcal{S}_B$ (namely $\textcolor{red}{1}$) determines the final decision in the tree.  The first constellation point to be sent is therefore $x_B = -9$, corresponding to the bit string $001\textcolor{red}{1}$.  He continues in this way, finally determining the symbols for transmission to be
\begin{equation}
\begin{aligned}
&001\textcolor{red}{1} \rightarrow -9 &= x_B(1) \\
&10\textcolor{red}{00} \rightarrow +1 &= x_B(2) \\
&110\textcolor{red}{1} \rightarrow +11 &= x_B(3) \\
\end{aligned}
\end{equation}
where $x_B(i)$ is sent during the $i^{th}$ time instance.
\end{example}

It is clear from the tree description that the secret bits encoded in $\mathcal{X}_B^{(k)}$ represent all bit strings of length $k$ uniformly.  Thus the assumption of uniformity in the random variables $P_B$ and $S_B$ which model Bob's public and secret input symbols guarantee that $X_B$ is uniform as well.  Moreover, it is clear from Proposition \ref{gent} that whenever Bob transmits $k$ secret bits (that is, transmits a constellation point $x_B\in \mathcal{X}_B^{(k)}$), Ray must guess uniformly at random from all possible bit strings of length $k$ when attempting to decode the last $k$ bits of the string corresponding to $x_B$.  Hence perfect secrecy is preserved, that is, $I(Y;S_B) = 0$.



\begin{theorem}\label{pam_rates}
The secrecy rates at Alice and Bob obtained using the present strategy are given by
\begin{equation}
\begin{aligned}
R^s_A &= m_A -3 + \frac{4}{M_A} \\
R^s_B &= 
m_A - \frac{4}{M_B}(M_A-1) 
\end{aligned}
\end{equation}
In particular, for fixed $M_A$ we have $\lim_{M_B\rightarrow\infty}R_B^s = m_A$.
\end{theorem}
\begin{IEEEproof}
This is a simple calculation given the sizes of the sets $\mathcal{X}_A^{(k)}$ and $\mathcal{X}_B^{(k)}$ in Proposition \ref{gent}.  In particular, since Alice transmits $k$ secret bits on $\mathcal{X}_A^{(k)}$, we have
\begin{equation}
R^s_A = \sum_{k = 1}^{m_A-2}k\frac{|\mathcal{X}_A^{(k)}|}{|\mathcal{X}_A|} = \frac{2}{M_A}\sum_{k = 1}^{m_A-2}k2^k.
\end{equation}
We now make use of the formula
\begin{equation}\label{useful}
\sum_{k = 1}^n k2^k = n2^{n+2} - (n+1)2^{n+1} + 2
\end{equation}
which can be easily proven by induction on $n$.  Plugging into the above completes the proof for the calculation of Alice's rate, and a similar computation is performed for Bob.
\end{IEEEproof}

\begin{figure}[t]
\centering
\includegraphics[angle=0,width=0.4\textwidth]{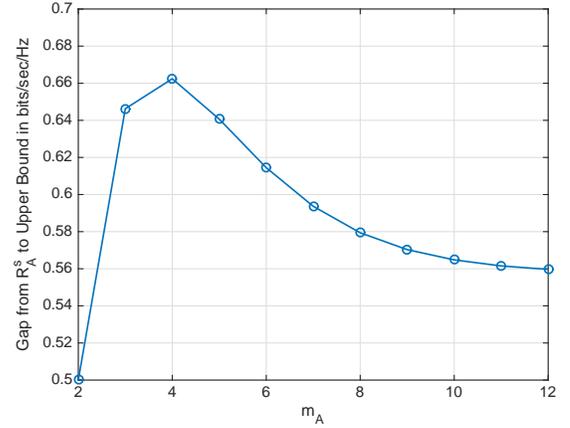}
\caption{The gap $\Delta_A = |\widetilde{R}^s_A-R^s_A|$ at Alice given the encoding scheme of Section \ref{sec:enc}, as a function of $m_A\geq 2$ and evaluated using the explicit formulas from Proposition \ref{explicit_gent} and Theorem \ref{pam_rates}.}\label{fig: alice_gap}
\end{figure}

\begin{figure}[t]
\centering
\includegraphics[angle=0,width=0.4\textwidth]{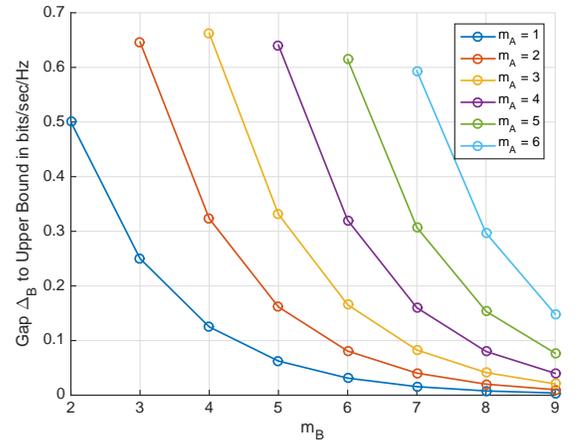}
\caption{The gap $\Delta_B = |\widetilde{R}^s_B-R^s_B|$ at Bob given the encoding scheme of Section \ref{sec:enc}, as a function of $m_B$ for various values of $M_B\geq 2M_A$ and evaluated using the explicit formulas from Proposition \ref{explicit_gent} and Theorem \ref{pam_rates}.}\label{fig: bob_gap}
\end{figure}

As the present scheme does not rely on any cooperation between Alice and Bob, we  compare the obtained secrecy rates to the corresponding upper bounds $\widetilde{R}^s_A$ and $\widetilde{R}^s_B$.  To measure the failure of our schemes to obtain the relevant upper bound, in Fig.\ \ref{fig: alice_gap} and Fig.\ \ref{fig: bob_gap} we plot the gaps
\begin{equation}
\Delta_A= |\widetilde{R}^s_A - R^s_A|,\quad \Delta_B = |\widetilde{R}^s_B - R^s_B|,
\end{equation}
respectively, as functions of $m_A$ and $m_B$ (recall that neither $\widetilde{R}^s_A$ nor $R^s_A$ depends on $m_B$, and that both the rate and the upper bound were zero when $m_A = 1$).

The gaps $\Delta_A$ and $\Delta_B$ are seen to be less than $0.7$ bits for all values of $M_A$ and $M_B$.  Furthermore, the gap $\Delta_B$ at Bob shrinks to less than $0.35$ bits for all $M_B\geq 4M_A$.  Note also that the gap $\Delta_B$ remains approximately constant for fixed $M_A/M_B$ and increasing $M_B$, while the rate $R^s_B$ and upper bound $\widetilde{R}^s_B$ increase without bound, thus showing that $R^s_B/\widetilde{R}^s_B\rightarrow 1$ as the sizes of the constellations increase.  These statements are in fact provable using the explicit expressions from Proposition \ref{explicit_gent} and Theorem \ref{pam_rates}, but as it involves a lengthy unintuitive calculation it will be omitted.  



\subsection{Explicit Encoder Construction with Cooperation}\label{sec:coop}

The scheme of the previous subsection left a large gap of approximately three bits between Alice's secrecy rate and the upper bound $\widehat{R}^s$, even asymptotically as the size of her constellation increases.  In this subsection we show that Alice's secrecy rate can greatly increase with knowledge of only $\lfloor s(x_B)\rfloor$ prior to transmission, where $\lfloor \cdot \rfloor$ is the floor function.  

Since we assume Alice has some knowledge of $x_B$, we judge our scheme against the upper bound $\widehat{R}^s$ of Theorem \ref{pam_cap}.  We remark that $\widehat{R}^s$ implicitly assumes Alice has full knowledge of $x_B$, while we assume much less and therefore this upper bound can be expected to be somewhat loose.

The assumption that Alice knows some information about Bob's signal is applicable if, for example, Bob transmits this information in a cryptographically secure manner to Alice before the current transmission cycle.  Equivalently, one could assume that Alice and Bob both know the value of a given jamming signal, of which Ray has no knowledge.


Knowing that $\lfloor s(x_B)\rfloor = k$ prior to transmission, Alice can potentially transmit $k$ secret bits in every $x_A\in \mathcal{X}_A$.  She exploits this knowledge by using the same tree-based bit labeling as in the previous section, but declares the last $k$ bits of every $x_A$ to be secret whenever $\lfloor s(x_B)\rfloor = k$.  The nature of the bit labeling guarantees that for a given $x_B$ satisfying $\lfloor s(x_B)\rfloor = k$, all bit strings of length $k$ are represented uniformly among Alice's symbols $x_A$.  Encoding public and secret bits proceeds as before.  

\begin{example}
Let us set $M_A = 8$ and $M_B = 32$.  For this example, we have
\begin{equation}
\begin{aligned}
\lfloor s(x_B)\rfloor &= 3 \Leftrightarrow x_B = \pm 1,\ldots, \pm 17 \\
\lfloor s(x_B)\rfloor &= 2 \Leftrightarrow x_B = \pm 19, \pm 21, \pm 23, \pm 25 \\
\lfloor s(x_B)\rfloor &= 1 \Leftrightarrow x_B = \pm 27, \pm 29 \\
\lfloor s(x_B)\rfloor &= 0 \Leftrightarrow x_B = \pm 31
\end{aligned}
\end{equation}
Suppose that Alice wishes to transmit the public and secret (respectively) bit strings
\begin{equation}
\mathcal{P}_A = 01,\quad \mathcal{S}_A = \textcolor{red}{1111011}
\end{equation}
and knows for the next three time instances that $\lfloor x_B(1)\rfloor = 3$, $\lfloor x_B(2)\rfloor = 2$, $\lfloor x_B(3)\rfloor = 2$.  She thus transmits $3$ secret bits in the first time instance, $2$ in the second, and $2$ in the third.  The output of the tree encoder is therefore
\begin{equation}
\begin{aligned}
\textcolor{red}{111}\rightarrow +7 &= x_A(1) \\
0\textcolor{red}{10}\rightarrow -3 &= x_A(2) \\
1\textcolor{red}{11} \rightarrow +7 &= x_A(3)
\end{aligned}
\end{equation}
Notice that for these values of $M_A$ and $M_B$, Alice's secrecy rate according to the scheme of the previous section was only $0.5$ bits/sec/Hz.
\end{example}

\begin{figure}[t]
\centering
\includegraphics[angle=0,width=0.4\textwidth]{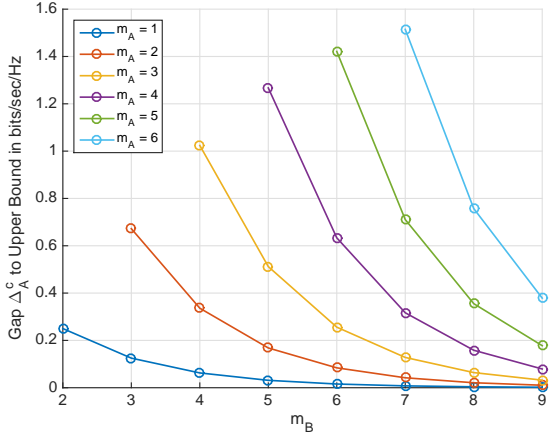}
\caption{The gap $\Delta_A^c =  |\widehat{R}^s - R^s_A|$ between Alice's secrecy rate given the encoding scheme of Section \ref{sec:coop} and the upper bound $\widehat{R}^s$, as a function of $M_B$ for various values of $M_A\leq 4M_B$.}\label{fig: alice_gap_coop}
\end{figure}


\begin{theorem}\label{alice_rates_coop}
Suppose that Alice and Bob employ $M_A$- and $M_B$-PAM modems with $M_B\geq 2M_A$, and assume that Alice has prior knowledge of the value of $\lfloor s(x_B)\rfloor$ for all $x_B$.  Then Alice can obtain the secrecy rate
\begin{equation}
R^s_A = m_A -\frac{4}{M_B}(M_A - 1) + \frac{2m_A}{M_B}.
\end{equation}
using the above-outlined encoder.  In particular, for fixed $M_A$ we have $\lim_{M_B\rightarrow \infty} R^s_A = m_A$.
\end{theorem}
\begin{IEEEproof}
Using the explicit formula for $|\psi^{-1}(y)|$ given in Proposition \ref{formula_for_psiy}, one can compute that
\begin{equation}
|\{x_B\ |\ \lfloor s(x_B)\rfloor = k\}| = \left\{\begin{array}{cl}
2^{k+1} & k \leq m_A -1 \\
M_B -2M_A + 2 & k = m_A
\end{array}\right. \nonumber
\end{equation}
Whenever Bob transmits $x_B$ such that $\lfloor s(x_B)\rfloor = k$, Alice transmits $k$ secret bits uniformly at random.  Her secrecy rate is therefore given by
\begin{equation}
R^s_A = m_A\frac{M_B -2M_A + 2}{M_B} + \sum_{k = 0}^{m_A-1}k\frac{2^{k+1}}{M_B}
\end{equation}
which straightforward simplification using (\ref{useful}) shows is equal to the quantity stated in the theorem.
\end{IEEEproof}

In Fig.\ \ref{fig: alice_gap_coop} we plot the gap $\Delta_A^c = |\widehat{R}^s - R^s_A|$ where $R^s_A$ is the secrecy rate at Alice obtained from Theorem \ref{alice_rates_coop}, and $\widehat{R}^s$ is the upper bound on the secrecy rates from Section \ref{sec_cap_explicit}.  While the gap $\Delta_A^c$ can be as much as $1.7$ bits for $M_B = 2M_A$, we note that one can prove $\Delta_A^c<0.9$ bits when $M_B = 4M_A$ and $\Delta_A^c<0.5$ bits for $M_B = 8M_A$.  Furthermore, while $\widehat{R}^s$ is a useful upper bound, it implicitly assumes Alice knows $x_B$ exactly rather than simply the value of $\lfloor s(x_B) \rfloor$, hence some non-trivial gap should be expected.  Lastly, we note that for a fixed ratio $M_A/M_B$, we see that $\Delta_A^c\rightarrow$ constant, demonstrating that $R^s_A/\widehat{R}^s\rightarrow 1$ as the sizes of the constellations increase.

\section{Generalization to the MIMO Relay Channel} \label{sec:MIMO}

The previous two sections have assumed that all noise sources are zero.  This section removes that assumption and simultaneously generalizes to a MIMO channel in which Alice and Bob each have $N$ antennas and Ray has $M\geq N$ antennas. We assume that Alice and Bob employ one of the encoders discussed in Section \ref{sec:enc}.  Having fixed a secrecy encoder, it is crucial to the success of Alice and Bob's transmission that Ray decode the sum $x_A + x_B$ of their symbols correctly, a point we now address by studying optimal precoding matrices at Alice and Bob.  

Let us denote the channel matrices Alice to Ray and Bob to Ray by $H_A$, $H_B\in \C^{M\times N}$, respectively, and the channel matrices from Ray to Alice and Ray to Bob, respectively, by $\tilde{H}_A$ and $\tilde{H}_B\in \C^{N\times M}$.  We assume all matrices are selected from a continuous distribution and are therefore full-rank with probability $1$.

Let $d\leq N$ and suppose Alice and Bob wish to transmit length $d$ vectors $x_A$ and $x_B$ of information symbols.  They employ linear precoders $G_A$, $G_B\in \C^{N\times d}$, so that Ray observes
\begin{equation}
y = H_AG_Ax_A + H_BG_Bx_B + w_R
\end{equation}
where $w_R$ is a length $M$ vector of additive noise, with entries i.i.d.\ zero-mean real Gaussian and variance $\sigma_R^2$ per real dimension.  To guarantee that Ray only observes sums of information symbols, Alice and Bob then must construct $G_A$ and $G_B$ to satisfy
\begin{equation}\label{secrecy_constraint}
H_AG_A = H_BG_B
\end{equation}
so that Ray observes
\begin{equation}
y = H_AG_A(x_A+x_B)+w_R
\end{equation}
from which he attempts to decode the sum $x_A+x_B$.  Hence condition (\ref{secrecy_constraint}) is crucial to employing the secrecy encoders of the previous section.

We further impose power constraints on Alice and Bob by first defining their average per-symbol power by
\begin{equation}\label{average_power}
P_A = \mathbf{E}|x_{A,i}|^2,\quad P_B = \mathbf{E}|x_{B,i}|^2
\end{equation}
where $x_A = (x_{A,1},\ldots,x_{A,d})^T$ and $x_{A,i}$ is a PAM symbol, and similarly for $x_{B,i}$.  To maintain this constraint after precoding we impose the conditions
\begin{equation}\label{power_constraint}
||G_A||^2_F \leq N,\quad ||G_B||^2_F\leq N.
\end{equation}
Here we recall that $||A||^2_F$ denotes the Frobenius norm of a matrix $A$, defined by $\sum_i \sigma_i^2$, where $\sigma_i$ are the singular values of $A$.  Note that assuming (\ref{secrecy_constraint}) and (\ref{average_power}), and after appropriate scaling of the channel matrices $H_A$ and $H_B$, we have that the average power of the received signal at Ray is $P_R = P_A + P_B$.

\subsection{Degrees of Freedom}

It is well-known and not hard to show (see \cite{ganesan,yener1,yener2}) that the \emph{degrees of freedom} $d$ of interference-free transmit dimensions available to both Alice and Bob is bounded by $d\leq (2N-M)^+$, and that every $d$ satisfying this inequality admits an interference-free transmission scheme.  Let us recall briefly how such schemes are constructed.  Let $\mathcal{H}_A$ and $\mathcal{H}_B$ be the column spans of $H_A$ and $H_B$, respectively, so that $\dim \mathcal{H}_A\cap \mathcal{H}_B = (2N-M)^+$.  If $d\leq (2N-M)^+$ then Alice and Bob decide on a $d$-dimensional subspace $\mathcal{S}$ of this intersection, and then choose $N\times d$ precoding matrices $G_A,G_B$ such that $\spanspace H_AG_A = \spanspace H_BG_B = \mathcal{S}$.  In a similar manner, $d\leq (2N-M)^+$ guarantees that Ray can transmit his PNC codewords back to Alice and Bob without the loss of any signal dimensions.

To ensure successful encoding and decoding which maximizes the degrees of freedom, we therefore restrict to $d$, $M$, and $N$ satisfying $1\leq d= 2N-M$ from now on.

\subsection{The Dimension of the Space of Precoders}

In this subsection we compute the dimension of the space of all $G_A$ and $G_B$ satisfying the secrecy constraint (\ref{secrecy_constraint}) and the power constraint (\ref{power_constraint}).  The dimension of this space measures the number of independent parameters when choosing precoding matrices, and determines the difficulty of optimizing the precoders numerically.  To be mathematically precise, `dimension' here means `dimension as a real manifold', but we omit the mathematical technicalities in favor of exposition.

If $G_A$ and $G_B$ are any precoders such that $||G_A||^2_F < N$ and $||G_B||^2_F < N$, then we can always improve the performance of the system by multiplying both precoders by a constant so that either $||G_A||^2_F = N$ or $||G_B||^2_F = N$.  So from now on we assume that one of the inequalities in (\ref{power_constraint}) is an equality.  On the other hand, the probability that both inequalities are actually equalities, e.g.\ that both precoders can be chosen to maximize both Alice and Bob's transmit power, is zero.

\begin{prop}\label{dim}
Fix $N\leq M$ and $d = 2N-M\geq1$, and let $H_A,H_B\in \C^{M\times N}$ be generic full-rank matrices.  Consider the matrix equation $H_AG_A = H_BG_B$ for some variable matrices $G_A,G_B\in \C^{N\times d}$ such that $||G_A||^2_F \leq N$ or $||G_B||^2_F \leq N$, and that exactly one of these inequalities is an equality.  Then
\begin{equation}
\dim_{\R}\mathcal{P} = 2(d^2-1).
\end{equation}
where $\dim_{\R}$ denotes dimension as a real manifold and $\mathcal{P}$ is the space of all such $G_A,G_B$ satisfying the above conditions.  
\end{prop}
\begin{IEEEproof}
As a real Euclidean space, the dimension of the space of all pairs $G_A,G_B\in \C^{N\times d}$ is $4Nd$.  Accounting for both real and imaginary parts, the equation $H_AG_A - H_BG_B = 0$ defines $2Md$ linear equations in the entries of $G_A,G_B$, all of which are independent by the assumptions that $H_A$ and $H_B$ are generic and full-rank.  Furthermore, suppose without loss of generality that $||G_A||^2_F = N$.  This single additional quadratic equation further reduces the dimension of the total space by two.  Putting this all together gives us $\dim_{\R}\mathcal{P} = 4Nd - 2Md - 2 = 2(d^2 - 1)$ as claimed.
\end{IEEEproof}

Thus when the degrees of freedom of the system is maximized, we have $\dim\mathcal{P} = 2(d^2-1)$ dimensions to optimize over when constructing optimal precoders.  When $(M,N) = (3,2)$ or $(5,3)$, for example, the dimension of $\mathcal{P}$ is zero and thus $\mathcal{P}$ consists of only isolated points, meaning that additional steepest descent optimization cannot improve system performance.  

\subsection{Optimizing Precoders at Alice and Bob}

While the previous section addressed the difficulty of optimizing precoders, in this subsection we address exactly what objective functions should be optimized.  Successful transmission between Alice and Bob requires Ray to accurately detect the sum $x_A + x_B$.  That is, assuming the secrecy constraint $H_AG_A = H_BG_B$ is satisfied, $G_A$ and $G_B$ should then be designed to maximize the mutual information at Ray.  We see that the task at hand is the following optimization problem:
\begin{equation}\label{opt_problem}
\begin{aligned}
 &\underset{G_A,G_B}{\text{maximize}}		&& I(Y;X_A+X_B) \\
 &\text{subject to} 		&& \left\{
 \begin{array}{c}
 \max\{||G_A||^2_F,||G_B||^2_F\}\leq N \\
 H_AG_B = H_BG_B
\end{array}
\right.
\end{aligned}
\end{equation}
where $Y = H_AG_AX_A + H_BG_BX_B + W_R$ as before.  Notice that the constraints exactly describe the space $\mathcal{P}$ of all precoders studied in Proposition \ref{dim}.  

It is important to note that the goal of the above optimization problem is not to increase the secrecy rate for either Alice or Bob, as the protocols of the previous section have already fixed this quantity.  Rather, we seek to increase the overall data rate of the total received signal at Ray, subject to the secrecy constraints.

\subsubsection{Zero-forcing precoders}  A straightforward zero-forcing scheme which satisfies the power and secrecy constraints of (\ref{opt_problem}) is the following.  Let $\begin{bmatrix} E_A \\ -E_B \end{bmatrix}$ be a $2N\times d$ matrix whose columns form a basis of the right nullspace of the $M\times 2N$ block matrix $[H_A\ H_B]$.  Now set
\begin{equation}\label{zf_precoders}
\begin{aligned}
G_A &= \sqrt{N}E_A/\gamma,\quad G_B = \sqrt{N}E_B/\gamma \\
\gamma &= \max\{||E_A||_F,||E_B||_F\}
\end{aligned}
\end{equation}
and it follows immediately that $H_AG_A = H_BG_B$ as desired.  When $N = M = d$, we can take  $G_A = \sqrt{N}H_A^{-1}/\gamma$ and $G_B = \sqrt{N}H_B^{-1}/\gamma$, where $\gamma = \max\{||H_A^{-1}||_F,||H_B^{-1}||_F\}$.

Every pair $G_A,G_B$ of precoding matrices satisfying the secrecy and power constraint can be constructed via the above process.  However, for certain parameters of $M$ and $N$, one can further optimize some initial zero-forcing scheme.

\subsubsection{Relaxation using the gap approximation}  As an attempt at an improvement on the above scheme, let us suppose that the power and secrecy constraints of (\ref{opt_problem}) are satisfied and use the gap approximation to approximate $I(Y;X_A+X_B)$ by the channel capacity:
\begin{equation}
C_{(H_A,H_B)} \approx I(Y;X_A+X_B) + \Gamma
\end{equation}
where $\Gamma$ is a constant.  By a well-known formula \cite{santipach}, the channel capacity (for fixed $H_A$, $H_B$) is then
\begin{equation}\label{gap_precoders}
C_{(H_A,H_B)} = \log_2\det\left(I_M + \frac{P_A+P_B}{\sigma_R^2}H_AG_AG_A^\dagger H_A^\dagger\right)
\end{equation}
The relaxation of the optimization problem at hand is then to maximize $C_{(H_A,H_B)}$ over all pairs $(G_A,G_B)$ subject to the same constraints of (\ref{opt_problem}). At first glance $H_B$ and $G_B$ are absent from this expression, but recall that we have already assumed that $H_AG_A = H_BG_B$.

To optimize (\ref{gap_precoders}) numerically, one performs steepest descent as follows.  Let $G = \begin{bmatrix} G_A \\ G_B \end{bmatrix}$, so that the task is to optimize over all possible $G$ satisfying the power and secrecy constraints.  The constraints of (\ref{opt_problem}) restrict the set of all possible $G$ to a bounded region of $\mathcal{N} = \nullspace\begin{bmatrix}H_A & -H_B\end{bmatrix}$.  One performs steepest descent on the coordinates of $G$ as normal, but after every iteration replaces $G$ with the projection $\text{Proj}_{\mathcal{N}}G$ and scales both blocks of $G$ to satisfy the power constraint of (\ref{opt_problem}).  We omit further details.

\begin{figure}[t]
\centering
\includegraphics[angle=0,width = 0.4\textwidth]{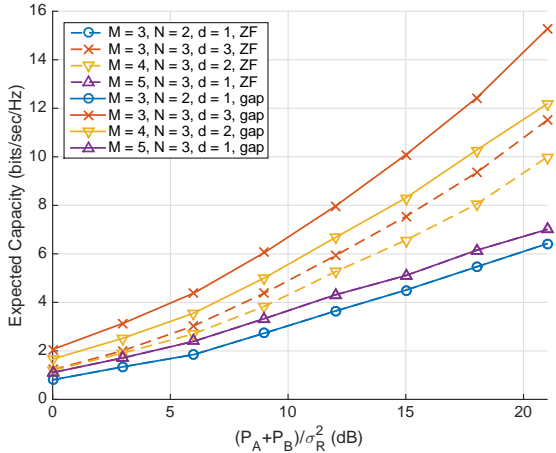}
\caption{The expected ergodic capacity $\mathbf{E}_{(H_A,H_B)}C_{(H_A,H_B)}$ as computed by (\ref{gap_precoders}) for randomly chosen zero-forcing precoders of (\ref{zf_precoders}), and those which have been further optimized using the gap approximation (\ref{gap_precoders}) of the mutual information $I(Y;X_A+X_B)$.} \label{cap_approx}
\end{figure}

In Fig.\ \ref{cap_approx} we plot the channel capacity (\ref{gap_precoders}).  Here the ``zero-forcing'' precoders were found according to (\ref{zf_precoders}).  The ``gap approximation'' precoders were then numerically optimized according to (\ref{gap_precoders}), and results were averaged over $10^3$ pairs $(H_A,H_B)$ of channel matrices.  The entries of $H_A$ and $H_B$ were drawn from i.i.d. complex zero-mean circularly symmetric Gaussian distributions with variance $1/M$.

The most notable feature of Fig.\ \ref{cap_approx} is that further optimization improves the precoding schemes for $(M,N) = (3,3)$ and $(4,3)$ quite a bit, but offers no improvement for the other two cases.  This is explained by Proposition \ref{dim}, since only for these parameters is $\dim_{\R}\mathcal{P} = d^2 -1 > 0$.  Secondly, the performance of the gap approximation precoder for $(M,N) = (4,3)$ is better than that of the zero-forcing precoder for $(M,N) = (3,3)$, even though the second scheme offers an additional degree of freedom.  Hence at the practical, finite SNR regimes of interest, specific precoding matrices may have more impact on capacity than the degrees of freedom.

\section{Conclusions and Future Work} \label{sec:Conclusions}
We have studied the potential for perfect secrecy in a single-relay network with two users.  Given finite, uniform input distributions, we have calculated upper bounds for the perfect secrecy rates under the assumptions that one user does or does not have information about the other user's signal, made these upper bounds explicit for PAM modems, and discussed the impact of synchronization errors on secrecy.  Two schemes that achieve perfect secrecy were presented using standard $M$-PAM modulators, one which assumes no cooperation between the users, and one which assumes the user with the smaller constellation knows some information about the other user's signal.  Gaps to the relevant upper bounds were shown to be small, especially asymptotically as the size of the larger constellation increases.  The system was generalized to a MIMO setting, and precoding matrices maintaining the required secrecy constraints were studied.  Finally, the potential for lattice encoders, alternative power allocation schemes, and applications to larger relay networks will be examined in the future.


\appendix

\subsection{Proof of Proposition 1}\label{app: Proof of Proposition 1}
\begin{IEEEproof}
We first show that the UBs in the noiseless scenario are given by
\begin{equation}\label{eq:SC noiseless Bob}
\widehat{R}^s_A = m_A - I(Y;X_A),\ \
\widehat{R}^s_B = m_B - I(Y;X_B). 
\end{equation}
where we recall that $m_A = \log_2 M_A$ and $m_B = \log_2 M_B$.  To see this note that in the noiseless scenario, the assumptions of Section \ref{sec:system model} guarantee that $I(Y_B;X_A|X_B) = m_A$, as Alice's symbol is perfectly recoverable given Bob's observation.  The above follows for Alice by substitution into (\ref{eq:SC Alice}), and the result for Bob follows by symmetry.

By the above it suffices to compute the mutual information $I(Y;X_A)$. We first fix a single $x_A\in\mathcal{X}_A$ and compute the marginal mutual information $I(Y;x_A)$.  An easy computation reveals that the joint distribution $(Y,X_A)$ has pmf
\begin{align}
p_{Y, X_A}(y,x_A) &= \frac{|\psi^{-1}(y)\cap \left(\{x_A\}\times \mathcal{X}_B\right)|}{M_AM_B} \nonumber\\
&= \left\{
\begin{array}{ll}
0, & \psi^{-1}(y)\cap\left(\left\{x_A\right\}\times \mathcal{X}_B\right) = \emptyset \\
\frac{1}{M_AM_B}, & \text{otherwise}
\end{array}
\right. \nonumber
\end{align}
Computing the mutual information $I(Y;x_A)$ now gives
\begin{align}
I(Y;x_A) &= \sum_{y\in\mathcal{Y}} p_{Y,X_A}(y,x_A)\log_2\left(\frac{p_{Y,X_A}(y,x_A)}{p_Y(y)p_{X_A}(x_A)}\right) \nonumber\\
&= \sum_{\substack{y\in\mathcal{Y} \\ y \in x_A + \mathcal{X}_B}}\frac{1}{M_AM_B}\log_2\frac{M_A}{|\psi^{-1}(y)|}
\end{align}
Summing up over all $x_A$, we arrive at
\begin{align}
I(Y;X_A) &= \sum_{x_A\in \mathcal{X}_A} I(Y;x_A) \nonumber\\
&= m_A - \sum_{x_A\in \mathcal{X}_A}\sum_{\substack{y\in\mathcal{Y} \\ y \in x_A + \mathcal{X}_B}}\frac{\log_2|\psi^{-1}(y)| }{M_AM_B}\nonumber\\
&= m_A - \sum_{y\in \mathcal{Y}}\log_2|\psi^{-1}(y)|\frac{|\psi^{-1}(y)|}{M_AM_B}
\end{align}
where the last equality follows by grouping like summands together.  An analogous calculation holds for Bob, and the proposition follows.
\end{IEEEproof}

\subsection{Proof of Theorem 1}\label{app: Proof of Theorem 1}
\begin{IEEEproof}
Define $S = \sum_{y\in \mathcal{Y}}\log_2|\psi^{-1}(y)|\cdot |\psi^{-1}(y)|$.  We can use the explicit formula for $|\psi^{-1}(y)|$ from Proposition \ref{formula_for_psiy} to write $S$ as $S = F+L$
where
\begin{equation}
\begin{aligned}
F &= \sum_{\substack{y = -(M_B-M_A) \\ y \text{ even}}}^{M_B-M_A}m_AM_A \\
L &= 2\sum_{\substack{y = M_B-M_A +2\\ y \text{ even}}}^{M_B+M_A-2} \log_2\left(\frac{M_B+M_A-y}{2}\right)\frac{M_B+M_A-y}{2} \nonumber
\end{aligned}
\end{equation}
The number of even integers in the interval $[-(M_B-M_A),M_B-M_A]$ is exactly $M_B-M_A+1$, and hence $F = m_AM_A(M_B-M_A+1)$.  Making the change of variables $a = \frac{M_B+M_A-y}{2}$ transforms the sum $L$ into $L = 2\sum_{a = 1}^{M_A-1}\log_2(a)a$.  We can conclude the proof by recalling that $\widehat{R}^s =S/M_AM_B = (F+L)/M_AM_B$, which is easily shown to be equal to the quantity in the theorem given the above calculations.
\end{IEEEproof}

\subsection{Proof of Proposition 4}\label{app:proof of prop 4}
\begin{IEEEproof}
By the symmetry of $\mathcal{X}_A$ we have $s(-x_A) = s(x_A)$.  When $x_A$ is positive it is clear that $|\psi^{-1}(x_A + x_B)|$ is minimized when $x_B = M_B -1$ and thus  
\begin{equation}
s(x_A) = s(|x_A|) = \log_2|\psi^{-1}(M_B-1+|x_A|)|
\end{equation}
Using the formula (\ref{formula_for_psiy}) for $|\psi^{-1}(y)|$ in the proof of Theorem \ref{pam_cap} and performing simple algebraic manipulations yields the following formula:
\begin{equation}
s(x_A) = \log_2\left(\frac{M_A + 1 - |x_A|}{2}\right)
\end{equation}
The first part of the proposition now follows easily from simple algebraic manipulations.  A similar calculation holds when computing $s(x_B)$.  Computing the sizes of the sets $\mathcal{X}_A^{(k)}$ and $\mathcal{X}_B^{(k)}$ follows straightforwardly from the definitions.
\end{IEEEproof}

\bibliographystyle{ieee}
\bibliography{myrefs_new}

\end{document}